\documentclass[twocolumn,epjc3]{svjour3}          % twocolumn

\RequirePackage[T1]{fontenc}

\smartqed  % flush right qed marks, e.g. at end of proof

\RequirePackage{graphicx}
\RequirePackage{mathptmx}      % use Times fonts if available on your TeX system
\RequirePackage{flushend}
\RequirePackage[numbers,sort&compress]{natbib}
\RequirePackage[colorlinks,citecolor=blue,urlcolor=blue,linkcolor=blue]{hyperref}

\journalname{Eur. Phys. J. C}

\begin{document}

\title{Excited State Mass spectra of doubly heavy baryons}

\subtitle{$\Omega_{cc}$, $\Omega_{bb}$ and $\Omega_{bc}$ }

\author{Zalak Shah\thanksref{addr1}
        \and
        Kaushal Thakkar\thanksref{addr2}
        and Ajay Kumar Rai\thanksref{e1,addr1}
        %and P. C. Vinodkumar\thanksref{addr3} %etc.
}

%\thankstext[$\star$]{t1}{Thanks to the title}
\thankstext{e1}{e-mail: raiajayk@gmail.com}
%\thankstext{e2}{e-mail: magic2@xxx.xx}

\institute{Department of Applied Physics, Sardar Vallabhbhai National Institute of Technology, Surat, Gujarat, India-395007\label{addr1}
          \and
 Department of Applied Sciences\& Humanities, GIDC Degree Engineering college, Abrama, Navsari, India-396406  \label{addr2}
          %\and
         % Department of physics, Sardar Patel University, V.V. Nagar, Anand, India-388120\label{addr3}
}

\date{Received: date / Accepted: date}

\maketitle

\begin{abstract}

We discuss the mass spectrum of $\Omega$ baryon with two heavy quarks and one light quark (\textit{ccs, bbs and bcs}). The main goal of the paper is to calculate the ground state masses and after that, the positive and negative parity excited states masses are also obtained  within a Hypercentral Constituent quark model, using coulomb plus linear potential framework. We also added first order correction to the potential. The mass spectra upto 5S for radial excited states and 1P-5P, 1D-4D and 1F-2F states for orbital excited states are computed for $\Omega_{cc}$, $\Omega_{bb}$ and $\Omega_{bc}$ baryons. Our obtained results are compared with other theoretical predictions which could be a useful complementary tool for the interpretation of experimentaly unknown heavy baryon spectra. The Regge trajectory is constructed in both ($n_r$, $M^{2}$) and ($J$, $M^{2}$) planes for $\Omega_{cc}$,$\Omega_{bb}$ and $\Omega_{bc}$ baryons and their slopes and intercepts are also determined. Magnetic moments of doubly heavy $\Omega'$s are also calculated.
\end{abstract}

\section{Introduction}

The doubly heavy $\Omega$ baryons represent a unique part of three-quark systems because they consist of strange light quark. Experiments and theoretical calculations have been diversely studied the heavy hadrons in last few years. So far singly heavy baryons have been discovered and quantum number of most of the observed states have been assigned. Many experiments, LHCb, BELLE, BABAR, FOCUS are planning to detect doubly and triply heavy baryons \cite{olive,2016}. The future project PANDA experiment at GSI are expected to give fruitful results in heavy baryon sector; especially charm sector. In fact, $\Xi_{cc}^{+}$ has been discovered \cite{selex} but none of the double/triple heavy $\Omega$ baryons have been discovered. Prediction for the masses of doubly heavy baryons have been presented by many authors so far. Indeed, this gives additional ground for new theoretical and experimental investigations of doubly as well as triply heavy baryonic properties. Many theoretical works have been focused on doubly heavy baryons, like relativistic three quark model \cite{1,ebert}, Salpeter model \cite{Giannuzzi}, heavy-quark effective theory \cite{korner}, QCD sum rule \cite{Jian,wang,aliev}, semiempirical mass formulas \cite{2}, the Hamiltonian model \cite{yoshida}, variational approach \cite{Roberts2008}, the three-body Faddeev method \cite{valcarce}, Hypercentral constituent quark model \cite{bhavin}, Lattice QCD \cite{mathur,paula,brown,can,alex,pacs}, etc. Recently, Ke-Wei Wei et al. used Regge phenomenology and with the quadratic mass relations, they obtained doubly and triply charmed baryon masses \cite{kwei}.
 
In this paper we shall study baryons containing two heavy quarks; charm-charm, bottom-bottom and charm-bottom with a light strange quark. Firstly, the mass spectra of these baryons ($\Omega_{cc}^{+}$, $\Omega_{bb}^{-}$ and $\Omega_{bc}^{0}$) are determined in the framework of Hypercentral constituent quark model (HCQM) \cite{bhavin, zalak,93,95,108}. In this paper we use coloumb plus linear potential and solve six dimensional hyperradial Schrodinger equation numerically. Secondly, the first order correction is taken into account in the potential. So, the purpose of our new investigation with an alternative calculation scheme of the baryon mass spectrum are elaborated as compared with the earlier performed investigations through the variational approach in Ref. \cite{bhavin}. The quantum numbers of doubly heavy ground state baryons are as folows:
\begin{itemize}
\item Strangeness S= -1  and Isospin I= 0 
\item spin-parity $J^{P}$=$\frac{1}{2}^{+}$ and $\frac{3}{2}^{+}$ 
\item Quark content \textit{ccs, bbs, bcs}
\end{itemize}

\noindent The calculations have performed for the radial excited states (upto 5S) and orbital excited states(1P-5P, 1D-4D and 1F-2F) at $\nu$=1.0\footnote{The different values of potential index $\nu$ from 0.5 to 2.0 had been taken during the calculations, but convenient results are obtained at 1.0.} for $\Omega_{cc}$, $\Omega_{bb}$ systems. To obtain the excited states of these doubly heavy baryons, we follow the same methodology used in previous calculations for singly charm baryons \cite{zalak}. Ground states are studied by various theoretical approaches for $\Omega_{cc}$, $\Omega_{bb}$ and $\Omega_{bc}$ are listed in Table~\ref{tab:1}.

The paper is organized as follows. After the introduction, we briefly describe our Hypercentral Constituent Quark Model in sec. 2. In Sec. 3 we present our results and drawn regge trajectories for $\Omega_{cc}$, $\Omega_{bb}$ and $\Omega_{bc}$ baryons. We also calculate magnetic moments for each baryons. In the last section, we drawn our conclusion.

\section{The HCQM Model}
\begin{table}
\begin{center}
\caption{The quark model parameters}
\label{tab:1}       % Give a unique label
\begin{tabular}{ll}
\hline\noalign{\smallskip}
Quark content&$m_{s}$ = 0.500 GeV\\ 
&$m_{c}$ = 1.275 GeV\\
&$m_{b}$ = 4.67 GeV\\
& \\
Model parameter&  $\tau = -\frac{2}{3} \alpha_{s}$\\
&$\alpha_{s}$=0.4155 GeV\\
&$C_{F}$=$\frac{2}{3}$ and $C_{A}$=3\\
\noalign{\smallskip}\hline
\end{tabular}
\end{center}
\end{table}

The baryons are made of three quarks and they are related to the Jacobi co-ordinates given as \cite{Bijker},
%\begin{subequations}
\begin{equation}
\vec{\rho} = \frac{1}{\sqrt{2}}(\vec{r_{1}} - \vec{r_{2}})
\end{equation}
\begin{equation}
\vec{\lambda} =\frac{m_1\vec{r_1}+m_2\vec{r_2}-(m_1+m_2)\vec{r_3}}{\sqrt{m_1^2+m_2^2+(m_1+m_2)^2}}
%\vec{\lambda}= \frac{1}{\sqrt{6}}(\vec{r_1}+ \vec{r_2}- 2\vec{r_3})
\end{equation}
%\end{subequations}
Here $m_i$ and $\vec{r_i}$ (i = 1, 2, 3) denote the mass and coordinate of the i-th constituent quark. Quarks masses are shown in Table~\ref{tab:1}. The respective reduced masses are given by
%\begin{subequations}
\begin{equation}
m_{\rho}=\frac{2 m_{1} m_{2}}{m_{1}+ m_{2}}
\end{equation}
\begin{equation}
 m_{\lambda}=\frac{2 m_{3} (m_{1}^2 + m_{2}^2+m_1m_2)}{(m_1+m_2)(m_{1}+ m_{2}+ m_{3})}
\end{equation}
The hyper radius $x= \sqrt{\rho^{2} + \lambda^{2}}$ and hyper angle $\xi= arctan (\frac{\rho}{\lambda})$ are in terms of the absolute values $\rho$ and $\lambda$ of the Jacobi coordinates \cite{130,131,132}. In the center of mass frame ($R_{c.m.} = 0$), the kinetic energy operator can be written as

\begin{equation}
-\frac{\hbar^2}{2m}(\bigtriangleup_{\rho} + \bigtriangleup_{\lambda})= -\frac{\hbar^2}{2m}\left(\frac{\partial^2}{\partial x^2}+\frac{5}{x}\frac{\partial}{\partial x}+\frac{L^2(\Omega)}{x^2}\right)
\end{equation}
where $L^2(\Omega)$=$L^2(\Omega_{\rho},\Omega_{\lambda},\xi)$ is the quadratic Casimir operator of the six-dimensional rotational group O(6) and its eigen functions are the hyperspherical harmonics, 

\[Y_{[\gamma]l_{\rho}l_{\lambda}}(\Omega_{\rho},\Omega_{\lambda},\xi)\]
satisfying the eigenvalue relation 

\[L^2Y_{[\gamma]l_{\rho}l_{\lambda}}(\Omega_{\rho},\Omega_{\lambda},\xi)=-\gamma (\gamma +4) Y_{[\gamma]l_{\rho}l_{\lambda}}(\Omega_{\rho},\Omega_{\lambda},\xi)\] 
Here, $\gamma$ is the grand angular momentum quantum number. In present paper, the confining three-body potential is chosen within a string-like picture, where the quarks are connected by gluonic strings and the potential strings increases linearly with a collective radius $r_{3q}$ as mentioned in \cite{ginnani2015}. Accordingly the effective two body interactions can be written as
\begin{equation}
\sum_{i<j}V(r_{ij})=V(x)+. . . .
\end{equation}

\begin{table}
\caption{Ground state masses of $\Omega_{cc}^{+}$,$\Omega_{bb}^{-}$ and $\Omega_{bc}^{0}$}
\label{tab:2}       % Give a unique label
\begin{tabular}{lll|ll|llll}
\hline\noalign{\smallskip}
Baryons& \multicolumn{2}{c}{$\Omega_{cc}^{+}$}&\multicolumn{2}{c}{$\Omega_{bb}^{-}$}&\multicolumn{2}{c}{$\Omega_{bc}^{0}$}\\
\hline
 $J^{P}$& $\frac{1}{2}^{+}$ &$\frac{3}{2}^{+}$ & $\frac{1}{2}^{+}$ &$\frac{3}{2}^{+}$ &$\frac{1}{2}^{+}$&$\frac{3}{2}^{+}$ \\
\hline
Our work&3.650&3.810&10.446&10.467&7.136&7.187\\
Ref. \cite{1}&3.719&3.746&10.422&10.432&6.999&7.024\\
Ref. \cite{ebert}&3.778&3.872&10.359&10.389&7.088&7.130\\
Ref. \cite{Giannuzzi}&3.648&3.770&10.271&10.289&6.994&7.017\\
Ref. \cite{Jian}&4.250&3.810&9.850&10.280&7.020&7.540\\
Ref. \cite{wang}&3.710&3.760&10.320&10.380&-&-\\
Ref. \cite{yoshida}&3.832&3.883&10.447&10.467&-&-\\
Ref. \cite{Roberts2008}&3.815&3.876&10.454&10.486&7.136&7.187\\
Ref. \cite{valcarce}&3.697&3.769&10.293&10.321&-&-\\
Ref. \cite{can}&3.747&3.819&-&-&-&-\\
Ref. \cite{paula}&3.713&3.785&-&-&-&-\\
Ref. \cite{brown}&3.738&3.822&10.273&10.308&6.999&7.059\\
Ref. \cite{alex}&3.654&3.724&-&-&-&-\\
Ref. \cite{kwei}&3.650&3.809&-&-&-&-\\
Ref. \cite{Ltang}&3.630&3.710&9.890&9.930&6.750&6.770\\
Ref. \cite{albertus}&3.702&3.783&10.260&10.297&6.986&7.046\\
Ref. \cite{zahra}&3.667&3758&10.397&10.495&7.103&7.200\\
Ref. \cite{5}&3.710&3.800&10.208&10.244&6.999&7.063\\
Ref. \cite{Kiselev2002}&3.590&3.690&10.180&10.200&6.910&6.990\\
Ref. \cite{Roncaglia}&3.740&3.820&10.370&10.400&7.045&7.120\\
\hline
\end{tabular}
\end{table}
In the hypercentral approximation, the potential is only depends on hyper radius(x). More details can be seen in references \cite{ginnani2015,M. Ferraris}. The hyper radius $x$ is a collective coordinate and therefore the hypercentral potential contains also the three-body effects. The  Hamiltonian of three body baryonic system in the hCQM is then expressed as
\begin{equation}
H=\frac{P_{x}^{2}}{2m} +V(x)
\end{equation}
where, $m=\frac{2 m_{\rho} m_{\lambda}}{m_{\rho} + m_{\lambda}}$, is the reduced mass and $x$ is the six dimensional radial hyper central coordinate of the three body system. The hyperradial Schrodinger equation corresponds to the above Hamiltonian can be written as,
\begin{equation}
\left[\frac{d^{2}}{d x^{2}} + \frac{5}{x} \frac{d}{dx} - \frac{\gamma(\gamma +4)}{x^{2}} \right] \Psi_{ \gamma}(x) = -2m[E- V(x)]\Psi_{ \gamma}(x)
\end{equation}
where $\Psi_{\gamma}$(x) is the hypercentral wave function and $\gamma$ is the grand angular quantum number. We consider a reduced hypercentral radial function, $\phi_{\gamma}(x) = x^{\frac{5}{2}}\Psi_{ \gamma}(x)$. Thus, six dimensional hyperradial Schrodinger equation reduces to,
\begin{equation}\label{eq:6}
\left[\frac{-1}{2m}\frac{d^{2}}{d x^{2}} + \frac{\frac{15}{4}+ \gamma(\gamma+4)}{2mx^{2}}+ V(x)\right]\phi_{ \gamma}(x)= E\phi_{\gamma}(x)
\end{equation}

\begin{table*}
\centering
\caption{Radial excited states masses of $\Omega_{cc}$, $\Omega_{bb}$ and $\Omega_{bc}$ Baryons (in GeV). A$\rightarrow$without first order correction and B$\rightarrow$with first order correction}
\label{tab:3}  
\begin{tabular}{llllllllll}
\hline\noalign{\smallskip}
Baryon&State & $J^{P}$ & A &B  &\cite{yoshida}&\cite{Roberts2008}&\cite{valcarce}&\cite{Giannuzzi}&\cite{ebert}\\
\noalign{\smallskip}\hline\noalign{\smallskip}

&2S	&$\frac{1}{2}^{+}$	&	4.028	&	4.041&4.227	&4.180&4.112&4.268&4.075\\
&	&$\frac{3}{2}^{+}$	&	4.085	&	4.096	&4.263&4.188&4.160&4.334&4.174	\\
$\Omega_{cc}$&3S	&	$\frac{1}{2}^{+}$	&	4.317	&	4.338	&4.295&&&4.714&4.321	\\
&	&$\frac{3}{2}^{+}$	&	4.345	&	4.365	&4.265&
&&4.776	\\
&4S	&$\frac{1}{2}^{+}$	&	4.570	&	4.598	&	\\
	&&$\frac{3}{2}^{+}$	&	4.586	&	4.614	&	\\
&5S	&$\frac{1}{2}^{+}$	&	4.801	&	4.836	&	\\
	&&$\frac{3}{2}^{+}$	&	4.811	&	4.845	&	\\
\noalign{\smallskip}\hline
&2S	&$\frac{1}{2}^{+}$	&	10.730	&	10.736	&10.707&10.693&10.604&10.830&10.610	\\
&&$\frac{3}{2}^{+}$	&	10.737	&	10.743	&10.723&10.721&10.622&10.839&10.645	\\
$\Omega_{bb}$&3S	&$\frac{1}{2}^{+}$&	10.973	&	10.983	&10.744&&&11.240&10.806	\\
&&$\frac{3}{2}^{+}$	&	10.976	&	10.986	&10.730&&&11.247&10.843	\\
&4S	&$\frac{1}{2}^{+}$&	11.191	&	11.205	&10.994	\\
	&&	$\frac{3}{2}^{+}$&	11.193	&	11.207	&	11.031\\
&5S	&$\frac{1}{2}^{+}$	&	11.393	&	11.411	&	\\
	&&$\frac{3}{2}^{+}$	&	11.394	&	11.412	&	\\
\noalign{\smallskip}\hline
&2S	&$\frac{1}{2}^{+}$	&	7.473	&	7.480	&		&&&	7.559	\\
&&$\frac{3}{2}^{+}$	&	7.490	&	7.497	&		&&&	7.571	\\
$\Omega_{bc}$&3S	&	$\frac{1}{2}^{+}$&	7.753	&	7.767	&		&&&	7.976	\\
	&&$\frac{3}{2}^{+}$	&	7.761	&	7.775	&		&&&	7.985	\\
&4S	&$\frac{1}{2}^{+}$	&	8.004	&	8.023	&		&\\
&&$\frac{3}{2}^{+}$&8.009	&	8.028	&		&\\
&5S&$\frac{1}{2}^{+}$&8.236	&	8.260	&		&\\
&&$\frac{3}{2}^{+}$&8.239	&	8.263	&		&\\
\noalign{\smallskip}\hline
\end{tabular}
\end{table*}

For the present study we consider the hypercentral potential V(x) as the color coulomb plus power potential with first order correction \cite{koma,11,20},
\begin{equation}\label{eq:7}
V(x) =  V^{0}(x) + \left(\frac{1}{m_{\rho}}+ \frac{1}{m_{\lambda}}\right) V^{(1)}(x)+V_{SD}(x)
\end{equation}
where $V^{0}(x)$ is given by
%\begin{subequations}
\begin{equation}
V^{(0)}(x)= \frac{\tau}{x}+ \beta x
\end{equation}
and first order correction as similar to the one given by \cite{koma},
\begin{equation}
V^{(1)}(x)= - C_{F}C_{A} \frac{\alpha_{s}^{2}}{4 x^{2}}
\end{equation}
%\end{subequations}
Here, $\tau$ is the hyper-Coulomb strength corresponds to the strong running coupling constant $\alpha_{s}$ \cite{zalak}. $\beta$ is the string tension of the confinement part of potential. $C_{F}$ and $C_{A}$ are the Casimir charges of the fundamental and adjoint representation. If we compare Eqn.(\ref{eq:6}) with the usual three dimensional radial Schrodinger equation, the resemblance between angular momentum and hyper angular momentum is given by \cite{9}, ${l(l+1)\rightarrow \frac{15}{4}+ \gamma(\gamma+4)}$. The spin-dependent part, $V_{SD}(x)$ of Eqn.(\ref{eq:7}) contains three types of the interaction terms \cite{12}. 
\begin{eqnarray}
%\label{eq:VSD}
V_{SD}(x)= V_{SS}(x)(\vec{S_{\rho}}.\vec{S_\lambda})
+ V_{\gamma S}(x) (\vec{\gamma} \cdot \vec{S})&&  \nonumber \\ + V_{T} (x)
\left[ S^2-\frac{3(\vec{S }\cdot \vec{x})(\vec{S} \cdot \vec{x})}{x^{2}} \right]
\end{eqnarray}

The spin-spin term $V_{SS} (x)$ gives the spin singlet triplet splittings, the spin-orbit term $V_{\gamma S}(x)$ and tensor term $V_{T}(x)$ describe the fine structure of the states. The detail of the terms are given in \cite{zalak}. We numerically solve the six dimensional Schrodinger equation using Mathematica notebook \cite{lucha}. We have followed the $^{(2S+1)} {\gamma}_{J}$ notations for spectra of baryons.

\begin{table*}
\centering
\caption{Orbitally excited states masses of $\Omega_{cc}$ Baryons (in GeV). A$\rightarrow$without first order correction and B$\rightarrow$with first order correction}
\label{tab:5}  
\begin{tabular}{lllllllll}
\noalign{\smallskip}\hline
State  & A &B  &\cite{yoshida}&\cite{Roberts2008}&\cite{paula}&\cite{ebert}&\cite{kwei}&Others\\
\noalign{\smallskip}\hline\noalign{\smallskip}
$(1^2P_{1/2})$&	3.964	&	3.989	&4.086&4.046&4.061&4.002&& 4.009\cite{valcarce}	\\
$(1^2P_{3/2})$&	3.948	&	3.972	&4.086 &4.052&4.132&4.102&3.910	\\
$(1^4P_{1/2})$&	3.972	&	3.998	&	\\
$(1^4P_{3/2})$&	3.956	&	3.981	&&&&&&3.960\cite{aliev}	\\
$(1^4P_{5/2})$&	3.935	&	3.958	&4.220 &4.152&&&4.058\\
\hline
$(2^2P_{1/2})$&	4.241	&	4.273	&4.199 &4.135&&4.251 &&4.101\cite{valcarce}	\\
$(2^2P_{3/2})$&	4.228	&	4.259	&4.201& 4.140&&4.345	\\
$(2^4P_{1/2})$&	4.248	&	4.280	&	\\
$(2^4P_{3/2})$&	4.234	&	4.266	&	\\
$(2^4P_{5/2})$&	4.216	&	4.247	&	\\
\hline
$(3^2P_{1/2})$&	4.492	&	4.529	&	\\
$(3^2P_{3/2})$&	4.479	&	4.517	&	\\
$(3^4P_{1/2})$&	4.498	&	4.536	&	\\
$(3^4P_{3/2})$&	4.486	&	4.523	&	\\
$(3^4P_{5/2})$&	4.469	&	4.506	&	\\
\hline
$(4^2P_{1/2})$&	4.723	&	4.767	&	\\
$(4^2P_{3/2})$&	4.712	&	4.755	&	\\
$(4^4P_{1/2})$&	4.728	&	4.772	&	\\
$(4^4P_{3/2})$&	4.717	&	4.761	&	\\
$(4^4P_{5/2})$&	4.703	&	4.745	&	\\
\hline
$(5^2P_{1/2})$&	4.939	&	4.989	&	\\
$(5^2P_{3/2})$&	4.929	&	4.978	&	\\
$(5^4P_{1/2})$&	4.944	&	4.994	&	\\
$(5^4P_{3/2})$&	4.934	&	4.984	&	\\
$(5^4P_{5/2})$&	4.921	&	4.969	&	\\
\hline
$(1^4D_{1/2})$&	4.156	&	4.186	&	\\
$(1^2D_{3/2})$&	4.133	&	4.162	&	\\
$(1^4D_{3/2})$&	4.141	&	4.170	&	\\
$(1^2D_{5/2})$&	4.113	&	4.141	&	4.264&4.202&&&4.153\\
$(1^4D_{5/2})$&	4.121	&	4.149	&	\\
$(1^4D_{7/2})$&	4.095	&	4.122	&&&&&4.294	\\
\hline
$(2^4D_{1/2})$&	4.407	&	4.446	&	\\
$(2^2D_{3/2})$&	4.389	&	4.425	&	\\
$(2^4D_{3/2})$&	4.395	&	4.432	&	\\
$(2^2D_{5/2})$&	4.372	&	4.407	&	\\
$(2^4D_{5/2})$&	4.378	&	4.414	&4.299&4.232	\\
$(2^4D_{7/2})$&	4.358	&	4.391	&	\\
\hline
$(3^4D_{1/2})$	&	4.446&	4.642	&	\\
$(3^2D_{3/2})$	&	4.425&	4.625	&	\\
$(3^4D_{3/2})$	&	4.432&	4.631	&	\\
$(3^2D_{5/2})$	&	4.407&	4.611	&	\\
$(3^4D_{5/2})$	&	4.414&	4.616	&4.410	\\
$(3^4D_{7/2})$	&	4.391&	4.598	&	\\
\hline
$(4^4D_{1/2})$&	4.863	&	4.911	&	\\
$(4^2D_{3/2})$&	4.847	&	4.894	&	\\
$(4^4D_{3/2})$&	4.853	&	4.900	&	\\
$(4^2D_{5/2})$&	4.833	&	4.879	&	\\
$(4^4D_{5/2})$&	4.838	&	4.885	&	\\
$(4^4D_{7/2})$&	4.821	&	4.866	&	\\
\hline
$(1^4F_{3/2})$&	4.313	&	4.348	&	\\
$(1^2F_{5/2})$&	4.287	&	4.321	&	\\
$(1^4F_{5/2})$&	4.294	&	4.328	&	\\
$(1^4F_{7/2})$&	4.271	&	4.303	&	\\
$(1^2F_{7/2})$&	4.264	&	4.296	&&&&& 4.383\\
$(1^4F_{9/2})$&	4.244	&	4.274	&&&&&4.516	\\
\hline
$(1^4F_{3/2})$&	4.552	&	4.593	&	\\
$(2^2F_{5/2})$&	4.530	&	4.569	&	\\
$(2^4F_{5/2})$&	4.536	&	4.575	&	\\
$(2^4F_{7/2})$&	4.515	&	4.553	&	\\
$(2^2F_{7/2})$&	4.509	&	4.547	&	\\
$(2^4F_{9/2})$&	4.490	&	4.527	&	\\
\noalign{\smallskip}\hline
\end{tabular}
\end{table*}

\begin{table*}
\centering
\caption{Orbitally excited states masses of $\Omega_{bb}$ Baryons (in GeV). A$\rightarrow$without first order correction and B$\rightarrow$with first order correction}
\label{tab:6}  
\begin{tabular}{llllllll}
\hline
State  & A &B  &\cite{yoshida}&\cite{Roberts2008}&\cite{ebert}&Others\\
\noalign{\smallskip}\hline\noalign{\smallskip}
$(1^2P_{1/2})$&10.634	&10.646	&10.607	&10.616&10.532	&10.519\cite{valcarce}\\
$(1^2P_{3/2})$&10.629&10.641&10.608	&10.619&10.566	\\
$(1^4P_{1/2})$&10.636&10.648		&		&	\\
$(1^4P_{3/2})$&10.631	&10.643&&		&&10.513\cite{aliev}	\\
$(1^4P_{5/2})$&10.625&10.637		&10.808&10.766	\\
\hline
$(2^2P_{1/2})$&10.881	&10.897	&10.796&10.763&10.738&10.683\cite{valcarce}	\\
$(2^2P_{3/2})$&10.877	&10.893		&10.797&10.765&10.775	\\
$(2^4P_{1/2})$&10.883	&10.899		&&&10.924	\\
$(2^4P_{3/2})$&10.879	&10.898		&&&10.961	\\
$(2^4P_{5/2})$&10.874	&10.888		&11.028	\\
\hline
$(3^2P_{1/2})$&	11.104&	11.124&	10.803&&11.083\\
$(3^2P_{3/2})$&	11.101&	11.120	&10.805	\\
$(3^4P_{1/2})$&	11.106&	11.125	&	\\
$(3^4P_{3/2})$&	11.103&	11.122	&	\\
$(3^4P_{5/2})$&	11.098&	11.177	&11.059	\\
\hline
$(4^2P_{1/2})$&11.310	&	11.332	&	\\
$(4^2P_{3/2})$&	11.307	&	11.339	&	\\
$(4^4P_{1/2})$&	11.312	&	11.334	&	\\
$(4^4P_{3/2})$&11.309	&	11.331	&	\\
$(4^4P_{5/2})$&	11.305	&		&	\\
\hline
$(5^2P_{1/2})$&	11.503	&11.528	&	\\
$(5^2P_{3/2})$&	11.500	&	11.525	&	\\
$(5^4P_{1/2})$&	11.504	&	11.530	&	\\
$(5^4P_{3/2})$&	11.502	&	11.527	&	\\
$(5^4P_{5/2})$&11.498	&	11.523	&	\\
\hline
$(1^4D_{1/2})$&	10.789	&	10.804	&	\\
$(1^2D_{3/2})$&	10.783	&	10.797	&	\\
$(1^4D_{3/2})$&	10.785	&	10.800	&	\\
$(1^2D_{5/2})$&	10.777	&	10.792	&10.729&10.720	\\
$(1^4D_{5/2})$&	10.779	&	10.794	&	\\
$(1^4D_{7/2})$&		&	10.786	&	\\
\hline
$(2^4D_{1/2})$&	11.017	&	11.036	&	\\
$(2^2D_{3/2})$&	11.012	&	11.030	&	\\
$(2^4D_{3/2})$&	11.014	&	11.032	&	\\
$(2^2D_{5/2})$&	11.008	&	11.025	&10.744&10.734	\\
$(2^4D_{5/2})$&	11.009	&	11.027	&	\\
$(2^4D_{7/2})$&	11.004	&	11.021	&	\\
\hline
$(3^4D_{1/2})$&	11.228	&	11.249	&	\\
$(3^2D_{3/2})$&	11.223	&	11.244	&	\\
$(3^4D_{3/2})$&	11.225	&	11.246	&	\\
$(3^2D_{5/2})$&	11.219	&	11.240	&10.937&	\\
$(3^4D_{5/2})$&	11.220	&	11.241	&	\\
$(3^4D_{7/2})$&	11.215	&	11.236	&	\\
\hline
$(4^4D_{1/2})$&	11.424	&	11.448	&	\\
$(4^2D_{3/2})$&	11.420	&	11.444	&	\\
$(4^4D_{3/2})$&	11.421	&	11.445	&	\\
$(4^2D_{5/2})$&	11.416	&	11.440	&	\\
$(4^4D_{5/2})$&	11.418	&	11.441	&	\\
$(4^4D_{7/2})$&	11.413	&	11.437	&	\\
\hline
$(1^4F_{3/2})$&	10.927	&	10.943	&	\\
$(1^2F_{5/2})$&	10.920	&	10.936	&	\\
$(1^4F_{5/2})$&	10.922	&	10.938	&	\\
$(1^4F_{7/2})$&	10.915	&	10.932	&	\\
$(1^2F_{7/2})$&	10.913	&	10.930	&	\\
$(1^4F_{9/2})$&	10.907	&	10.924	&	\\
\hline
$(2^4F_{3/2})$&	11.142	&	11.162	&	\\
$(2^2F_{5/2})$&	11.136	&11.155		&	\\
$(2^4F_{5/2})$&	11.137	&11.157		&	\\
$(2^4F_{7/2})$&	11.132	&11.151		&	\\
$(2^2F_{7/2})$&	11.130	&11.149		&	\\
$(2^4F_{9/2})$&	11.125	&11.144		&	\\
\noalign{\smallskip}\hline
\end{tabular}
\end{table*}

\section{Results and Disscussions}

\subsection{Mass Spectra}

The ground and excited states of doubly heavy $\Omega$ baryons are still experimentally unknown to us. Therefore, we have calculated the ground as well as excited state masses of doubly heavy baryons $\Omega_{cc}$, $\Omega_{bb}$ and $\Omega_{bc}$[See Table~\ref{tab:2}-~\ref{tab:8}].  Such kind of theoretical study is very much useful to obtain their experimental states ($J^{P}$ values), masses and other properties. These mass spectra of doubly heavy baryons are obtained by using coulomb plus linear potential in Hypercentral constituent quark model. Our  computed ground states with $J^{P}=\frac{1}{2}^{+}$ and $\frac{3}{2}^{+}$ are compared with different theoretical approaches in Table~\ref{tab:2}. Our estimated ground state masses of $\Omega_{cc}^{+}$ have difference in range of $\approx$ 100MeV with other predictions, whereas $\Omega_{bb}^{-}$ and $\Omega_{bc}^{0}$ masses are higher than others.

\noindent The Lattice QCD calculations for excited states of $\Omega_{cc}$ has been performed by Padmanath et al. upto $\frac{7}{2}^{+}$ and $\frac{7}{2}^{-}$ parities \cite{mathur1}. They use variational approach in which they try to write down the eigen states in terms of the operators and determine the energies from the evolution of the correlators of the eigenstates. Energy splittings of the $\Omega_{cc}$ states are from the mass of the $\eta_c$ meson. Ref. \cite{zhi} estimate the ground state mass of $\Omega_{cc}$ to be around 3.726 GeV in chiral perturbation theory.

The radial excited state masses for these three baryons are computed from 2S-5S and are compared with Refs. \cite{yoshida, Roberts2008, valcarce, Giannuzzi,ebert} in Table~\ref{tab:3}. We can observe that, our 2S and 3S states show less MeV difference with ref. \cite{ebert} than other references for $\Omega_{cc}^{+}$ and $\Omega_{bb}^{-}$. Next, in case of $\Omega_{bc}$ only ref. \cite{Giannuzzi} has calculated radial excited states upto 3S. We noticed, our 2S and 3S state masses (A(B)) with $J^{P}$=$\frac{1}{2}^{+}$ are 76(69) and 225(211), while 80(73) and 224(210) (with $J^{P}$=$\frac{3}{2}^{+}$) lower respectively. Note that, A are masses without first order correction and B are masses by adding first order correction in Table[~\ref{tab:3}-~\ref{tab:8}].

The orbital excited states are calculated for 1P-5P, 1D-4D and 1F-2F mentioned in Table~\ref{tab:5}-~\ref{tab:8}. Isospin splittings were also considered, that means we have considered all possible combination of total spin S and total angular momentum J to obtain orbital states mass spectra. We can observe that the total combinations for P, D and F states are 5, 6 and 6 respectively. Other theoretical approaches have also been calculated these orbital excited states but they have not consider all (S, J) combinations. We have compared our outcomes with other models in following tables \ref{tab:5}-~\ref{tab:8}. 

Our obtained orbital excited masses(A) are compared and discussed with other predictions in following paragraph. For $\Omega_{cc}$,
our 1P state $J^{P}=\frac{1}{2}^{-}$ show 38 MeV(with \cite{ebert}),  $J^{P}=\frac{3}{2}^{-}$ show 38 MeV and $J^{P}=\frac{5}{2}^{-}$ show 123 MeV (with \cite{kwei}) difference. Our 2P state $J^{P}=\frac{1}{2}^{-}$ show 10 MeV(with \cite{ebert}),  $J^{P}=\frac{3}{2}^{-}$ show 27 MeV (with \cite{yoshida}) difference. Our 1D state $J^{P}=\frac{5}{2}^{+}$ show 40 MeV and $J^{P}=\frac{3}{2}^{-}$ show 199 MeV difference with \cite{kwei}. In 2D-3D states difference is 79 MeV and 4 MeV for $J^{P}=\frac{5}{2}^{+}$ with ref. \cite{yoshida}. Our 1F state values are 119($J^{P}=\frac{7}{2}^{-}$) and 272($J^{P}=\frac{9}{2}^{-}$) MeV lower than ref. \cite{kwei}. 

\noindent For $\Omega_{bb}$, our 1P state $J^{P}=\frac{1}{2}^{-}$ and $J^{P}=\frac{3}{2}^{-}$ are 18 and 10 MeV higher than ref. \cite{Roberts2008}. Ref. \cite{yoshida} masses for $J^{P}=\frac{1}{2}^{-}$, $J^{P}=\frac{3}{2}^{-}$ and $J^{P}=\frac{5}{2}^{-}$ are 85, 80 and 154 MeV lower than our prediction. Our 3P state show 39 MeV difference with \cite{yoshida}. Our 1D state $J^{P}=\frac{5}{2}^{+}$ show 48 MeV and 57 MeV difference with \cite{yoshida} and \cite{Roberts2008} respectively. The orbital mass spectra of third doubly heavy baryon, $\Omega_{bc}$ is not calculated in any approach. Refs. \cite{Kiselev2002} and \cite{ebert} stated that excited levels are not possible for $\Omega_{bc}$ because the excited states of diquarks $\lbrace bc\rbrace$ are not stable due to the emission of soft gluons. We have not consider di-quark mechanism in our approach, so that, we calculated orbital mass spectra for $\Omega_{bc}$ baryon. We guess, we are the first to calculate these spectra. 
\begin{table}
\centering
\caption{Orbitally excited states masses of $\Omega_{bc}$ Baryons (in GeV). A$\rightarrow$without first order correction and B$\rightarrow$with first order correction}
\label{tab:8}  
\begin{tabular}{llll}
\hline
State  & A &B  \\
\hline
$(1^2P_{1/2})$&	7.375	&	7.386	&	\\
$(1^2P_{3/2})$&	7.363	&	7.373	&	\\
$(1^4P_{1/2})$&	7.381	&	7.392	&	\\
$(1^4P_{3/2})$&	7.369	&	7.379	&	\\
$(1^4P_{5/2})$&	7.353	&	7.363	&	\\
\hline
$(2^2P_{1/2})$&	7.657	&	7.674	&	\\
$(2^2P_{3/2})$&	7.647	&	7.664	&	\\
$(2^4P_{1/2})$&	7.662	&	7.679	&	\\
$(2^4P_{3/2})$&	7.652	&	7.669	&	\\
$(2^4P_{5/2})$&	7.639	&	7.655	&	\\
\hline
$(3^2P_{1/2})$&	7.912	&	7.935	&	\\
$(3^2P_{3/2})$&	7.903	&	7.925	&	\\
$(3^4P_{1/2})$&	7.916	&	7.939	&	\\
$(3^4P_{3/2})$&	7.908	&	7.930	&	\\
$(3^4P_{5/2})$&	7.896	&	7.918	&	\\
\hline
$(4^2P_{1/2})$&	8.147	&	8.175	&	\\
$(4^2P_{3/2})$&	8.140	&	8.167	&	\\
$(4^4P_{1/2})$&	8.151	&	8.179	&	\\
$(4^4P_{3/2})$&	8.143	&	8.171	&	\\
$(4^4P_{3/2})$&	8.133	&	8.160	&	\\
\hline
$(5^2P_{1/2})$&	8.368	&	8.400	&	\\
$(5^2P_{3/2})$&	8.361	&	8.393	&	\\
$(5^4P_{1/2})$&	8.372	&	8.404	&	\\
$(5^4P_{3/2})$&	8.365	&	8.396	&	\\
$(5^4P_{5/2})$&	8.355	&	8.386	&	\\
\hline
$(1^2D_{1/2})$&	7.562	&	7.577	&	\\
$(1^2D_{3/2})$&	7.545	&	7.561	&	\\
$(1^4D_{3/2})$&	7.551	&	7.566	&	\\
$(1^2D_{5/2})$&	7.531	&	7.547	&	\\
$(1^4D_{5/2})$&	7.536	&	7.552	&	\\
$(1^4D_{7/2})$&	7.518	&	7.534	&	\\
\hline
$(2^2D_{1/2})$&	7.821	&	7.843	&	\\
$(2^2D_{3/2})$&	7.807	&	7.829	&	\\
$(2^4D_{3/2})$&	7.812	&	7.834	&	\\
$(2^2D_{5/2})$&	7.795	&	7.816	&	\\
$(2^4D_{5/2})$&	7.799	&	7.821	&	\\
$(2^4D_{7/2})$&	7.784	&	7.805	&	\\
\hline
$(3^2D_{1/2})$&	8.060	&	8.088	&	\\
$(3^2D_{3/2})$&	8.048	&	8.075	&	\\
$(3^4D_{3/2})$&	8.052	&	8.079	&	\\
$(3^2D_{5/2})$&	8.037	&	8.063	&	\\
$(3^4D_{5/2})$&	8.041	&	8.068	&	\\
$(3^4D_{7/2})$&	8.028	&	8.054	&	\\
\hline
$(4^2D_{1/2})$&	8.285	&	8.317	&	\\
$(4^2D_{3/2})$&	8.274	&	8.305	&	\\
$(4^4D_{3/2})$&	8.277	&	8.309	&	\\
$(4^2D_{5/2})$&	8.263	&	8.294	&	\\
$(4^4D_{5/2})$&	8.267	&	8.298	&	\\
$(4^4D_{7/2})$&	8.254	&	8.285	&	\\
\hline
$(1^4F_{3/2})$&	7.721	&	7.742	&	\\
$(1^2F_{5/2})$&	7.702	&	7.723	&	\\
$(1^4F_{5/2})$&	7.708	&	7.728	&	\\
$(1^4F_{7/2})$&	7.690	&	7.711	&	\\
$(1^2F_{7/2})$&	7.685	&	7.705	&	\\
$(1^4F_{9/2})$&	7.670	&	7.690	&	\\
\hline
$(2^4F_{3/2})$&	7.721	&	7.965	&	\\
$(2^2F_{5/2})$&	7.702	&	7.949	&	\\
$(2^4F_{5/2})$&	7.708	&	7.953	&	\\
$(2^4F_{7/2})$&	7.690	&	7.938	&	\\
$(2^2F_{7/2})$&	7.685	&	7.934	&	\\
$(2^4F_{9/2})$&	7.670	&	7.921	&	\\
\noalign{\smallskip}\hline
\end{tabular}
\end{table}
\subsection{Regge Trajectory}
\begin{figure}
\centering
\resizebox{0.55\textwidth}{!}{%
  \includegraphics{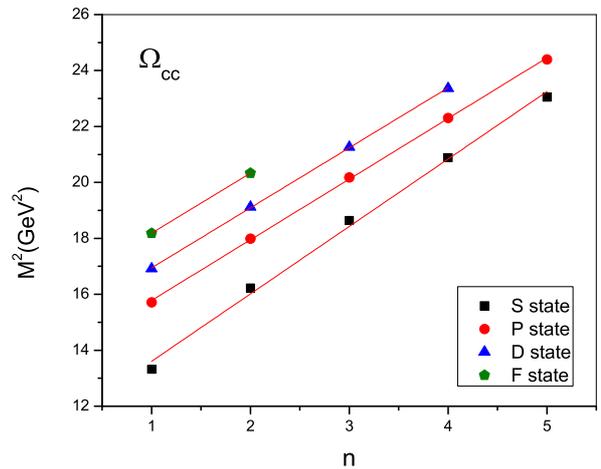}
}
\caption{Regge Trajectory ($M^{2}$ $\rightarrow$ n) for $\Omega_{cc}$ baryon.}
\label{fig:1}       % Give a unique label
\end{figure}

\begin{figure}
\centering
\resizebox{0.55\textwidth}{!}{%
  \includegraphics{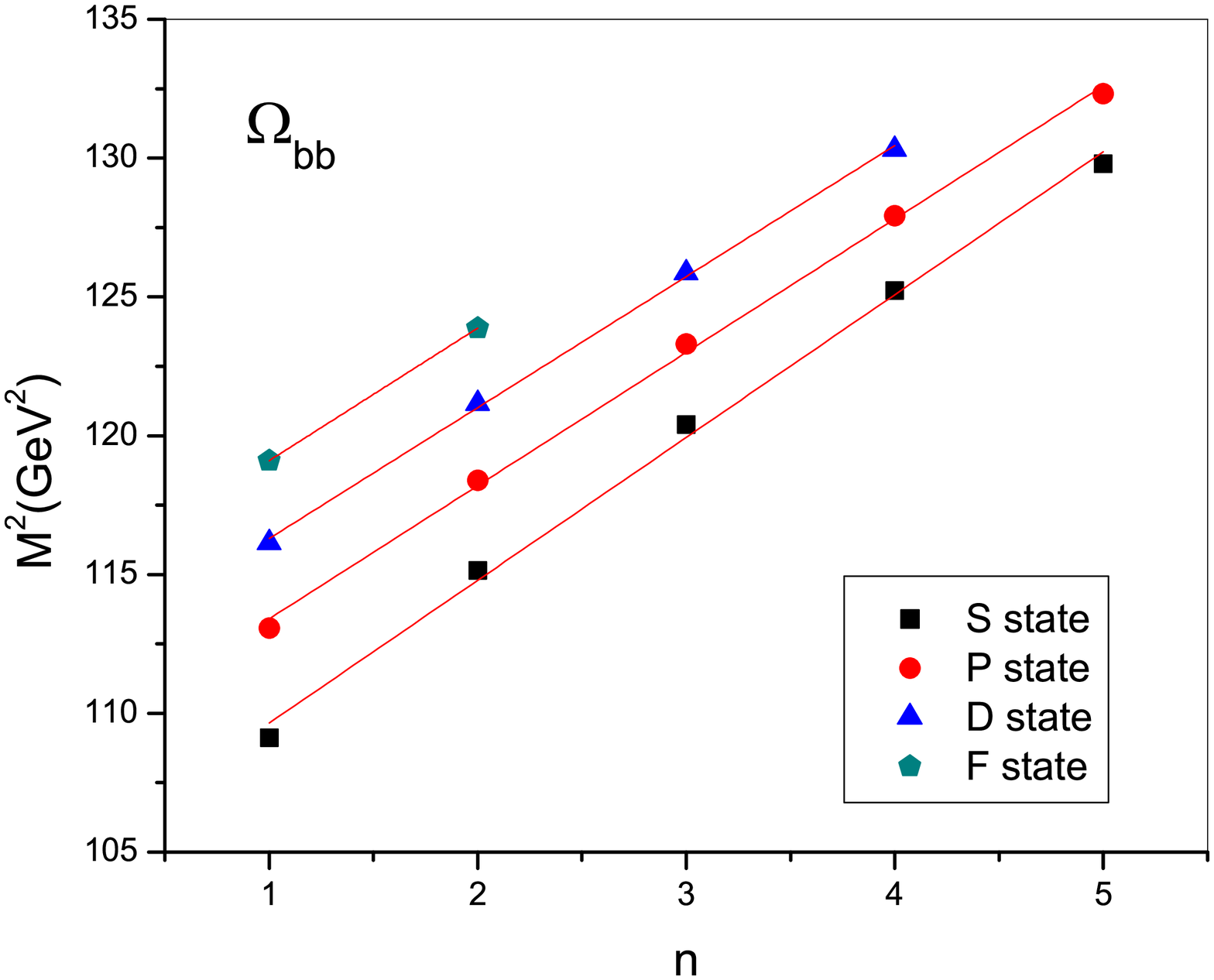}
}
\caption{Regge Trajectory ($M^{2}$ $\rightarrow$ n) for $\Omega_{bb}$ baryon.}
\label{fig:2}       % Give a unique label
\end{figure}

\begin{figure}
\centering
\resizebox{0.55\textwidth}{!}{%
  \includegraphics{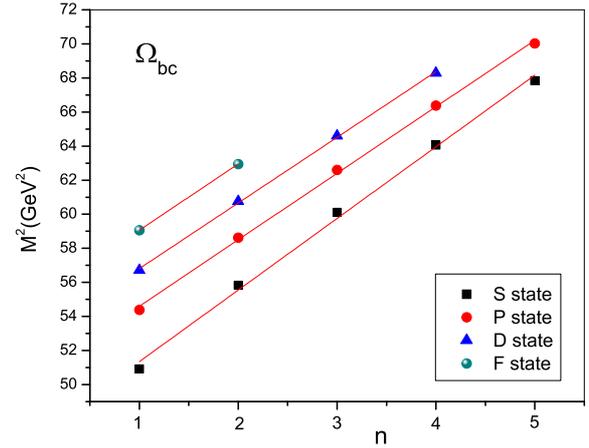}
}
\caption{Regge Trajectory ($M^{2}$ $\rightarrow$ n) for $\Omega_{bc}$ baryon.}
\label{fig:3}       % Give a unique label
\end{figure}

\begin{figure*}
\centering
\resizebox{1.0\textwidth}{!}{%
  \includegraphics{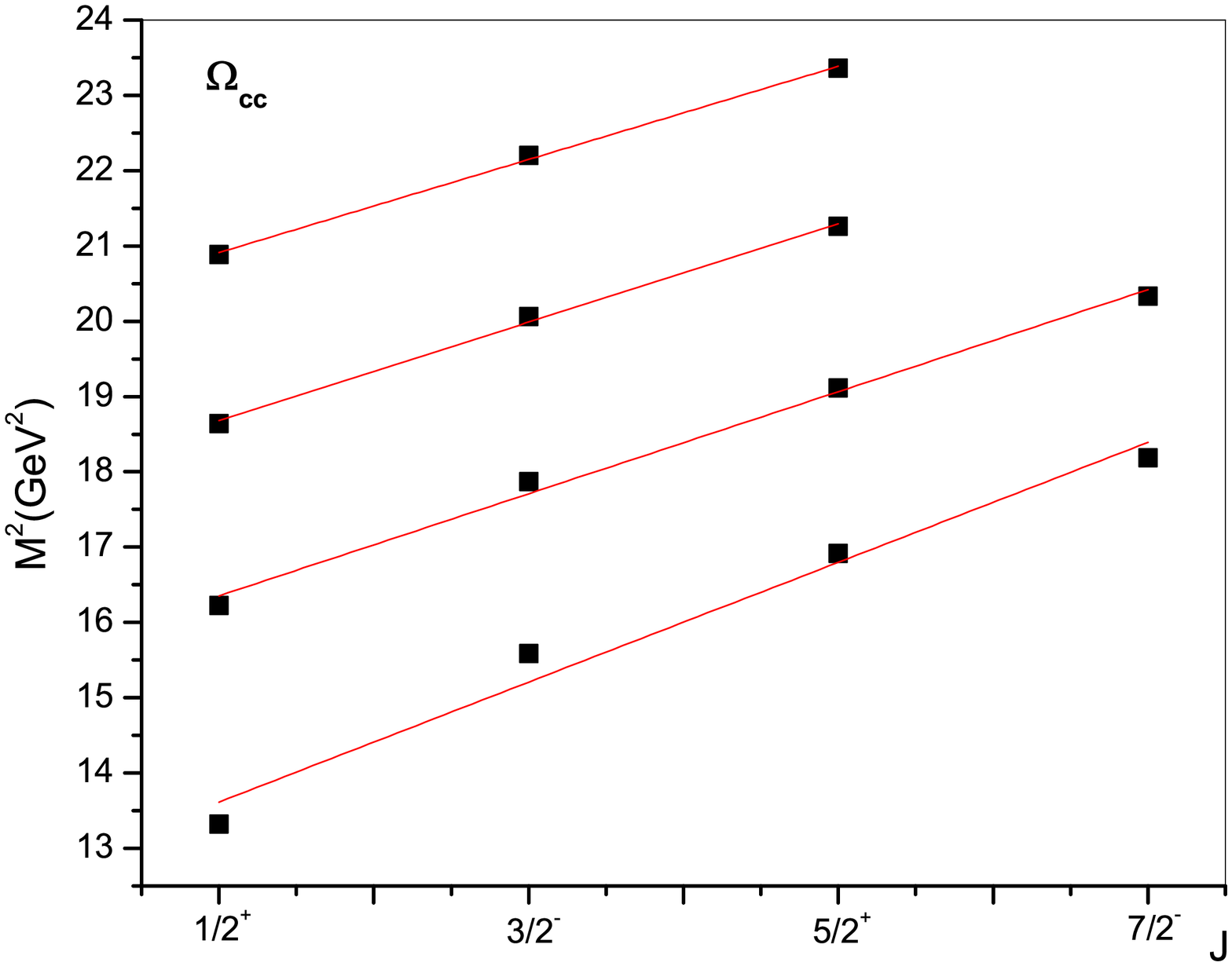}
  \includegraphics{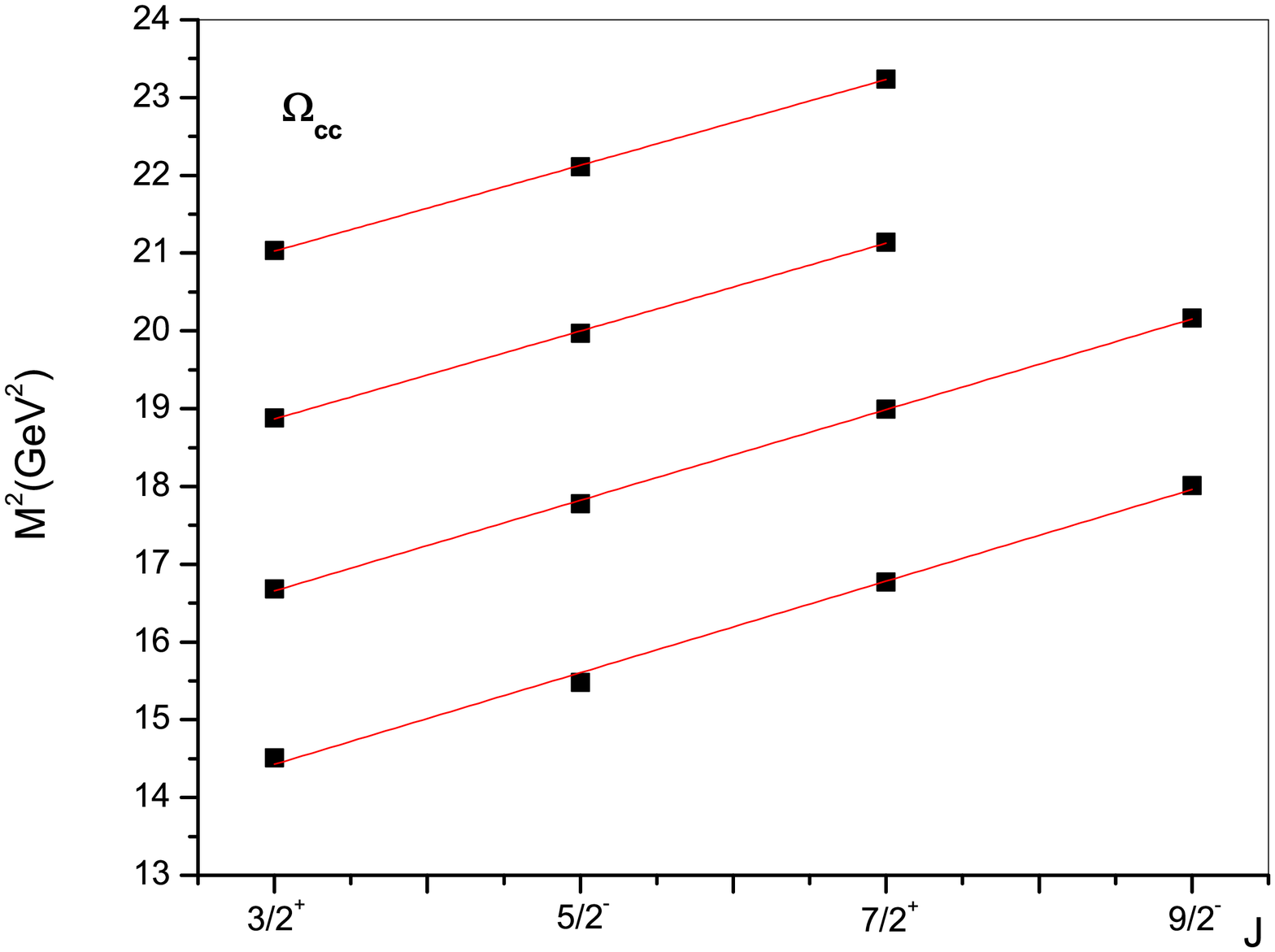}
}
\caption{Regge Trajectory ($M^{2}$ $\rightarrow$ J) for $\Omega_{cc}$ baryon.}
\label{fig:4}       % Give a unique label
\end{figure*}

\begin{figure*}
\centering
\resizebox{1.0\textwidth}{!}{%
  \includegraphics{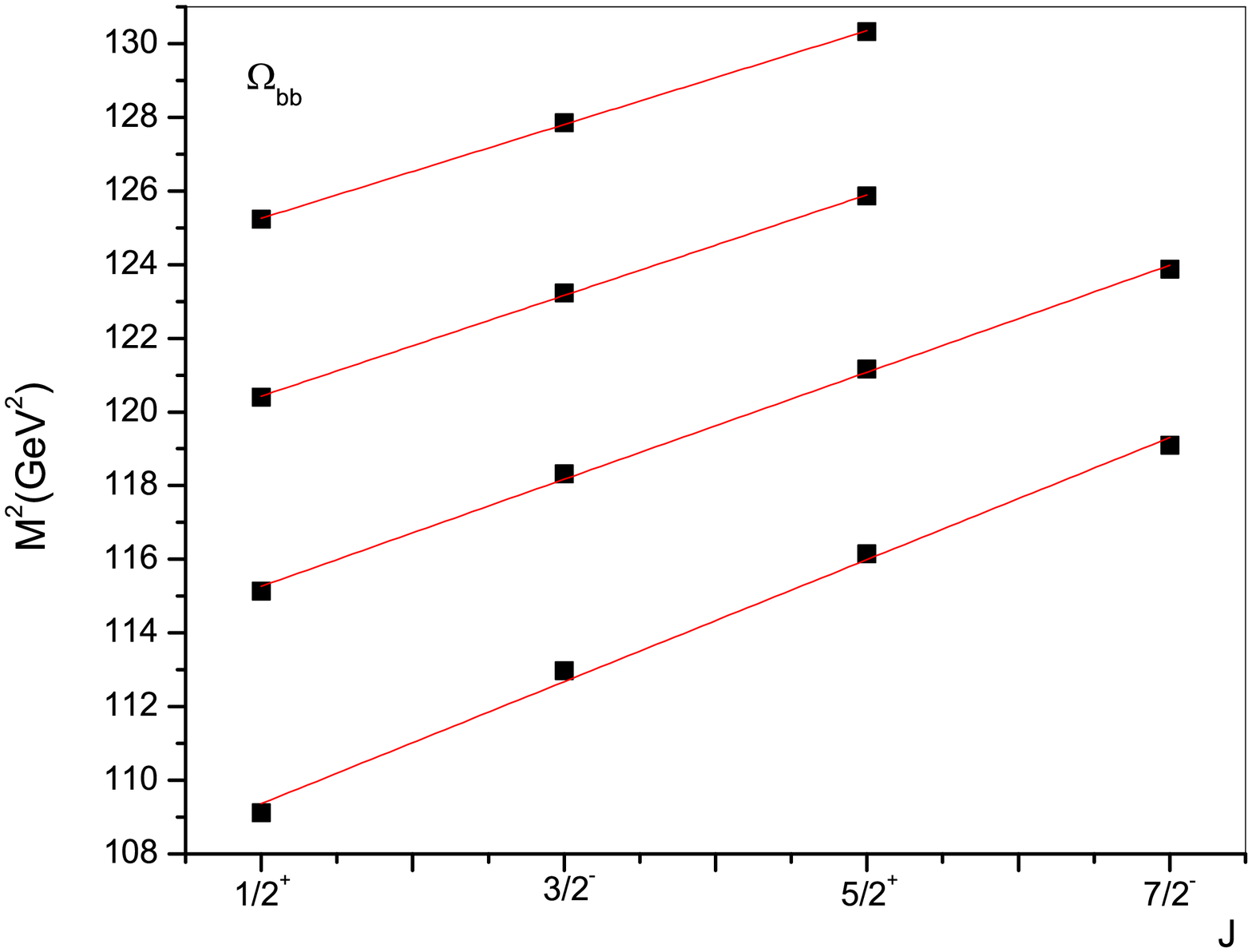}

  \includegraphics{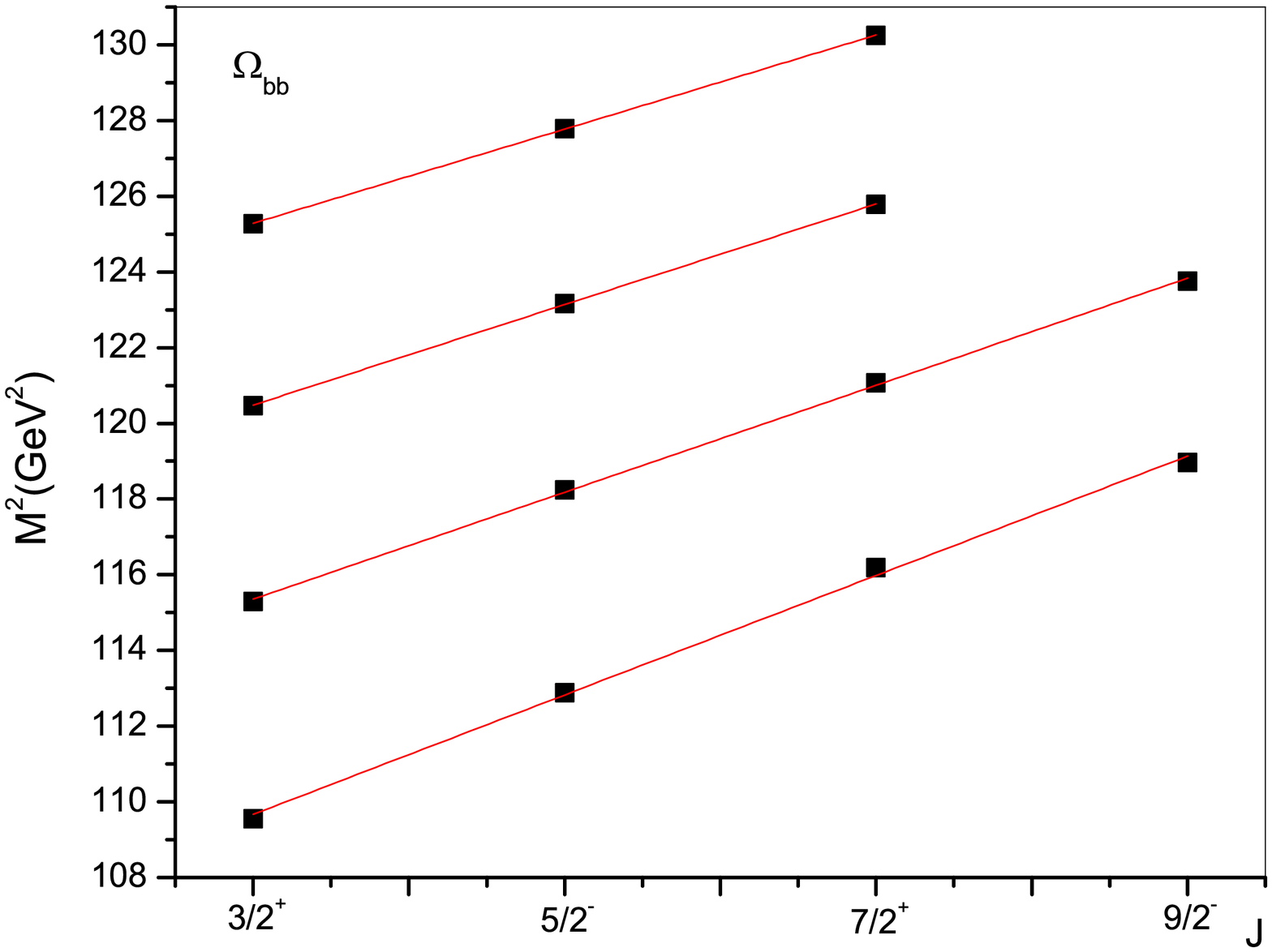}
}
\caption{Regge Trajectory ($M^{2}$ $\rightarrow$ J) for $\Omega_{bb}$ baryon.}
\label{fig:5}       % Give a unique label
\end{figure*}

We calculated both radial and orbital excited states masses upto L=3. Using them we are able to construct Regge trajectories in (n, $M^{2}$) and (J, $M^{2}$) planes. n is principal quantum number and J is a total spin. The Regge trajectories are presented in Figs.~\ref{fig:1}-~\ref{fig:5} . Straight lines were obtained by the linear fitting in all figures. The ground and radial excited states S ($J^{P}=\frac{1}{2}^{+}$) and the orbital excited state P ($J^{P}= \frac{1}{2}^{-}$), D ($J^{P}= \frac{5}{2}^{+}$) and F ($J^{P}= \frac{7}{2}^{-}$) are plotted in Figs~\ref{fig:1}-~\ref{fig:3} from bottom to top. We use,
 \begin{equation}
n_r=\beta M^2+ \beta_{0}
\end{equation}
\noindent Where, $\beta$ and $\beta_{0}$ are slope and intercept, respectively and $n_{r}$= n-1.  The fitted slopes and intercepts are given in Table~\ref{tab:7}. 

\noindent We use natural($J^{P}=\frac{1}{2}^{+}$, $J^{P}=\frac{3}{2}^{-}$, $J^{P}=\frac{5}{2}^{+}$, $J^{P}=\frac{7}{2}^{-}$) and unnatural ($J^{P}=\frac{3}{2}^{+}$, $J^{P}=\frac{5}{2}^{-}$, $J^{P}=\frac{7}{2}^{+}$, $J^{P}=\frac{9}{2}^{-}$) parity masses and plotted graphs for $\Omega_{cc}$ and $\Omega_{bb}$ [See Fig. ~\ref{fig:4}-~\ref{fig:5}]. For that we use,
 \begin{equation}
J=\alpha M^2+ \alpha_{0}
\end{equation}
\noindent Where, $\alpha$ and $\alpha_{0}$ are slope and intercept, respectively. The fitted slopes and intercepts for both naural and unatural parities are given in Table~\ref{tab:11}.
We observe that the square of the calculated masses fit very well to the linear trajectory and almost parallel and equidistant in S, P, D and F states. We can determine the possible quantum numbers and prescribe them to particular Regge trajectory with the help of our obtained results.
\begin{table}
\centering
\caption{Fitted slopes and intercepts of the regge trajectories in (n, $M^{2}$) plane.}
\label{tab:7}
\begin{tabular*}{\columnwidth}{@{\extracolsep{\fill}}lllllll@{}}
\hline
 Baryon&$J^{P}$ &State& $\beta$&$\beta_{0}$\\
%\multicolumn{2}{@{}l}{parameter} & Set 1 & Set 2\\
\hline
$\Omega_{cc}$&$\frac{1}{2}^{+}$&S&0.413$\pm$0.0144&-5.614$\pm$0.270\\
&$\frac{1}{2}^{-}$&P&0.461$\pm$0.004 &-7.273$\pm$ 0.09\\
&$\frac{3}{2}^{-}$&P&0.460$\pm$0.05
&-7.195$\pm$0.091\\
&$\frac{5}{2}^{+}$&D&0.466$\pm$0.003&-7.890$\pm$0.007\\
\hline
$\Omega_{bb}$&$\frac{1}{2}^{+}$&S&0.194
$\pm$0.006&-21.242$\pm$0.752\\
&$\frac{1}{2}^{-}$&P&0.208$\pm$0.004 &-23.583
$\pm$0.560\\
&$\frac{3}{2}^{-}$&P&0.208
$\pm$0.05&-23.520$\pm$0.056\\
&$\frac{5}{2}^{+}$&D&0.212$\pm$0.004&-24.613$\pm$0.490\\
\hline
$\Omega_{bc}$&$\frac{1}{2}^{+}$&S&0.194
$\pm$0.006&-21.242$\pm$0.752\\
&$\frac{1}{2}^{-}$&P&0.255$\pm$0.004 &-13.885
$\pm$0.282\\
&$\frac{3}{2}^{-}$&P&0.208
$\pm$0.05&-23.520$\pm$0.056\\
&$\frac{5}{2}^{+}$&D&0.259$\pm$0.004&-14.731$\pm$0.238\\
\hline
\end{tabular*}
\end{table}

\begin{table*}
\centering
\caption{Fitted parameters $\alpha$ and $\alpha_0$ are slope and Intercept of parent and daughter Regge trajectories. Columns 3,4 are for natural parities and 5,6 are for unnatural parities.}
\label{tab:11}
\begin{tabular}{cccccccccccc}
\hline
 Baryon&Trajectory& $\alpha$&$\alpha_{0}$& $\alpha$&$\alpha_{0}$\\
%\multicolumn{2}{@{}l}{parameter} & Set 1 & Set 2\\
\hline
$\Omega_{cc}$&parent&0.614$\pm$0.0653&-7.332$\pm$1.051&0.846$\pm$0.0358
&-11.202$\pm$0.582\\
&1 daughter&0.733$\pm$0.039&-10.983$\pm$0.731&0.858$\pm$0.0125&-13.291$\pm$0.231\\
&2 daughter&0.762$\pm$0.039&-13.225$\pm$0.783&0.884$\pm$0.017&-15.688$\pm$0.341\\
&3 daughter&0.807$\pm$0.0295&-15.897$\pm$0.655&0.906$\pm$0.0108&-180.055$\pm$0.241\\
\hline
$\Omega_{bb}$&parent&0.301$\pm$0.013&-31.881$\pm$1.523&0.316$\pm$0.009&-33.699$\pm$1.081\\
&1 daughter&0.344$\pm$0.009&-38.61$\pm$1.052&0.354$\pm$0.005&-39.804$\pm$0.649\\
&2 daughter&0.366$\pm$0.008&-43.065$\pm$0.955&0.376$\pm$0.004&-44.354$\pm$0.437\\
&3 daughter&0.393$\pm$0.006&-48.172$\pm$0.794&0.402$\pm$0.003&-49.381$\pm$0.341\\
\hline
\end{tabular}
\end{table*}

\subsection{Magnetic moments}
The magnetic moment of baryons are obtained in terms of the spin, charge and effective mass of the bound quarks as \cite{bhavin, 93,95,108}

\begin{eqnarray}\nonumber
\mu_{B}=\sum_{i}\langle \phi_{sf}\vert \mu_{iz}\vert\phi_{sf}\rangle)
\end{eqnarray}
where
\begin{equation}
\mu_{i}=\frac{e_i \sigma_i}{2m_{i}^{eff}}
\end{equation}
$e_i$ is a charge and $\sigma_i$ is the spin of the respective constituent quark corresponds to the spin flavor wave function of the baryonic state.  The effective mass for each of the constituting quark $m_{i}^{eff}$ can be defined as
\begin{equation}
m_{i}^{eff}= m_i\left( 1+ \frac{\langle H \rangle}{\sum_{i} m_i} \right)
\end{equation}
where, $\langle H \rangle$ = E + $\langle V_{spin} \rangle$. Using these equations, we calculated magnetic moments of $\Omega_{cc}^{+}$, $\Omega_{bb}^{-}$ and $\Omega_{bc}^{-}$ baryons. The spin flavor wave function \cite{107} and magnetic moments are given in Table {~\ref{tab:table10}. Our obtained ground state magnetic moments are also compared with others and results are reasonably close.
\begin{table*}
\centering
\caption{\label{tab:table10} Magnetic Moment(in nuclear magnetons) of $J^{P}$~ $\frac{1}{2}^{+}$ and $\frac{3}{2}^{+}$ doubly heavy baryons.}
\begin{tabular}{cccccccccccc}
\hline
Baryons& wave-function&Our&\cite{103}&\cite{bhavin}&\cite{106}&\cite{101}&\cite{109}&\cite{iv}\\
\hline
  $\Omega_{cc}^{+}$& $\frac{4}{3}\mu_{c}$- $\frac{1}{3}\mu_{s}$&0.692&0.668&0.785&0.635&-&0.639&0.66\\
 $\Omega_{cc}^{+ *}$& $\mu_{c}$+$\mu_{s}$&0.285&0.332&0.121&0.139&0.210&-\\
 $\Omega_{bb}^{-}$&$\frac{4}{3}\mu_{b}$- $\frac{1}{3}\mu_{s}$&0.108&0.120&0.109&0.101&0.111&0.100\\
$\Omega_{bb}^{- *}$&2$\mu_{b}$+$\mu_{s}$&-1.239&-0.730&-0.711&-0.662&-0.703&-\\
 $\Omega_{bc}^{0}$&$\frac{2}{3}\mu_{b}+\frac{2}{3}\mu_{c}$- $\frac{1}{3}\mu_{s}$&0.439&0.034&0.397&0.368&0.399&-\\
  $\Omega_{bc}^{0 *}$&$\mu_{b}$+$\mu_{c}$+$\mu_{s}$&-0.181&-0.111&-0.317&-0.261&-0.274&-\\
  \hline
\end{tabular}
\end{table*}

\section{Conclusions}

The ground state as well as excited state masses are obtained for doubly heavy $\Omega$ baryons in hCQM are tabulated in Table~\ref{tab:2}-~\ref{tab:8}. We have compared our results with other approaches and they are not in the mutual agreement with each other. Moreover, all are unknown experimentally.  Due to that, we can not single out any model. It is very important to compare results with lattice QCD but most of them have calculated only ground state masses of these baryons \cite{brown, pacs, alex}. We have also constructed the Regge trajectories from obtained masses. This study will help future experiments to identify the baryonic states from resonances. In addition, we have calculate the magnetic moments and they are fairly agreement with other calculations. We have successfully employed this model to doubly heavy $\Omega$ baryons. Now, we would like to calculate the properties of $\Xi$ baryons in future.

\begin{acknowledgements}
Z. Shah wants to thanks M. Padmanath for providing their data of Lattice QCD. A. K. Rai acknowledges the financial support extended by DST, India  under SERB fast track scheme SR/FTP /PS-152/2012. We are very much thankful to Prof. P. C. Vinodkumar for his valuable suggestions throughout the work.
\end{acknowledgements}

\end{document}